\documentclass[12pt]{article}
\usepackage{graphicx}
\usepackage{cite}
\usepackage{amsmath}
\usepackage{color}
\usepackage{hyperref}

\definecolor{darkblue}{rgb}{0,0,.7}
\definecolor{darkred}{rgb}{0.7,0,0}

\hypersetup{colorlinks=true, linkcolor=black,citecolor=black,urlcolor=black}

\numberwithin{equation}{section}

\newcommand{\TEXTsymbol}[1]{$#1$}

\title{A Technical Review\ of Penning Trap based Investigations in Neutron
Decay}

\author{J. Byrne\\
\small Department of Physics and Astronomy,\\
\small University of Sussex,\\
\small Brighton, Sussex BN1 9QH}

\begin{document}

\maketitle

\begin{abstract}
This review is concerned with a detailed analysis of some of the technical
problems which arise in the application of the Penning trap method to the
experimental study of neutron $\beta $-decay, a technique which was first
successfully tested on the low-flux swimming-pool reactor LIDO (capture flux
= 3$\cdot 10^6$cm$^{-2}$s$^{-1})$ at AERE, Harwell in the 1970's. It does
not discuss the scientific merits or demerits of these studies. Of
particular importance are the trapping and release of neutron decay protons,
and the influence of magnetic mirror effects and radial drifting on the
trapped particles. Since these have energies \TEXTsymbol{<}1 keV they must
be accelerated to energies of order 20-30 keV following release, at which
point they are recorded in a silicon surface barrier detector. However
serious difficulties were encountered in the post-release acceleration
process associated with vacuum breakdown in the presence of crossed electric
and magnetic fields.

\end{abstract}

\textbf{\#1 The Role of the Neutron Lifetime in Astrophysics and Particle
Physics.}

The availability of a precise value for the lifetime of the free neutron is
of major importance in astrophysics because it is this quantity which
ultimately determines the rate at which hydrogen is transmuted into helium
by thermonuclear processes in the sun$^{\text{ }[1]}$. According to big-bang
scenarios the neutron lifetime also influences the rate of primordial helium
production, but in this case persistent disagreements between cosmologists
and nuclear experimentalists have been satisfactorily resolved$^{\text{ }[2%
]} $. Thus the continuing interest in neutron lifetime measurements centres
on the crucial role this number plays in fixing precise values for the weak
coupling constants in beta-decay$^{\text{ }[3]}$, and for arriving at a
nuclear-structure-independent value for the $V_{ud}$ element in the CKM
quark mixing matrix $^{[4-5]}$.

The important relationship connecting the lifetime $\tau _n=t/\ln (2),$
where $t$ is the half-life, with the vector and axial vector weak coupling
constants $G_V$ and $G_A$ is given by 
\begin{equation}
\mathcal{F}t=\left[ 2\pi ^3\left( \ln 2\right) \hbar ^7/m_e^5c^4\right]
/\left( G_V^2+3G_A^2\right) =K/(G_V^2\left[ 1+3\mid \lambda \mid ^2\right]
),\,  \tag{1.1}
\end{equation}
where $\lambda =G_A/G_V$, $K/\left( \hbar c\right) ^6=\left( 8.1202787\pm
0.000011\right) \cdot 10^{-7}$GeV$^{-4}s$ and $\mathcal{F}=1.71489\pm
0.000002$ $^{[6]}$ is the integrated Fermi phase-space factor including
model-dependent and model-independent radiative corrections.

Neutron lifetime experiments are notoriously difficult, and this is so for
essentially three reasons: (i) neutron decay is a rare process which is
difficult to isolate against an intense background of $\gamma $-rays from
nuclear interactions, (ii) absolute neutron counting relies for its
precision on an array of physical and chemical data, e.g. cross sections,
surface densities, isotopic ratios etc., (iii) absolute counting of the
charged decay products requires a detailed understanding of the
electromagnetic forces to which these particles are subjected during
transport from the source volume to the detector. Many of these difficulties
are avoided in stored neutron experiments, in which ultra-cold neutrons are
confined in suitable magnetic field configurations$^{\text{ }[7-8]}$, or in
material bottles$^{\text{ }[9-10]}$. However these techniques have their own
sources of systematic error which have not as yet been entirely clarified.
In this communication we confine attention to the technical aspects of the
Penning trap method for studying neutron decay, and examine in some detail
some of the technical problems which have arisen in successive versions of
the method$^{\text{ }[11-14]}$.\medskip

\textbf{\# 2} \textbf{The Penning Trap Method.}

The operating principle of the Penning trap is based on the Penning
cold-cathode vacuum gauge $^{[15]}$ and the description ''Penning trap'' was
introduced by Dehmelt $^{[16]}$ in his study of the electron g-factor
anomaly. The Penning trap has proved to be an extremely versatile instrument
in fundamental physics$^{\text{ }[17]}$, and its application to the neutron
lifetime problem goes back some four decades $^{[11]}$. In this method
protons from neutron decay, which have energies less than about 0.75 keV,
are stored in a Penning trap before being ejected and counted in a silicon
surface barrier counter, maintained at a \textit{negative} potential of
20-30 kV. This technique has the double advantage that the source volume can
be precisely determined, and that the background is reduced in the ratio of
detection time to storage time. In the ideal Penning trap an axially
symmetric electrostatic quadrupole potential is superimposed on a coaxial
uniform magnetic field, a combination in which the trapped charged particles
undergo harmonic oscillations along the axis and epicycloidal motion in the
transverse plane$^{\text{ }[18]}$. However these ideal conditions are
unnecessarily restrictive in a trap for protons from neutron decay, where
there is no requirement for the axial motion to be harmonic. Thus suitable
quasi-Penning traps may be formed in a wide range of axially symmetric
electric field configurations, based on the two-cylinder electrostatic lens$%
^{\text{ }[19]}.$This system of proton trap and detector functions when a
minimal set of conditions is met. In particular it is essential to have (a)
a magnetic field of sufficient strength that the radius of cyclotron motion
is small in comparison with the dimensions of the apparatus, (b) an
electrostatic potential well approximately 1kV in depth, (c) a fast negative
pulse to open the trap, and (d) a negative accelerating potential $>20kV$.

The adiabatic invariants associated with motion in the ideal Penning trap
may be calculated by application of the Hamilton Jacobi equation$^{\text{ }[%
20]}.$Evaluation of the adiabatic invariants associated with the $\phi $-
and z-coordinates goes ahead in a simple manner and we find 
\begin{equation}
(a)\;J_\phi =\oint p_\phi \,d\phi =\pi m\omega _r\,(a^2-R^2)=J_a-J_R\text{ ;}%
\quad (b)\;J_z=\oint p_z\,dz=\pi m\omega _z\,Z^2  \tag{2.1}
\end{equation}
where $p_\phi $ , $p_r$ and $p_z$ are canonical momenta, $a$ is the radius
of the cyclotron orbit, $R$ is the radial coordinate of the guiding centre
and $Z$ is the amplitude of the axial oscillation. For the r-coordinate the
situation is slightly more complicated in that the two cases: (a) a%
\TEXTsymbol{>}R (cyclic accelerator), and (b) a\TEXTsymbol{<}R (Penning trap)%
$,$ must be treated separately. In the latter case we find then that 
\begin{equation}
J_r=\oint p_r\,dr=\pi m\omega _rR^2=J_R  \tag{2.2}
\end{equation}
Since, assuming strict cylindrical symmetry, 
\begin{equation}
p_\phi =mr^2(\dot{\phi}-\omega _c/2)  \tag{2.3}
\end{equation}
is a constant of the motion, it follows that $J_a$ is also an adiabatic
invariant.

Both the quantities J$_a$ and J$_R$ have simple physical interpretations.
Thus $eJ_a/2\pi $ is the magnetic moment of the cyclotron orbit traced out
by the particle circulating in the magnetic field, while $J_R$ is
proportional to the magnetic flux linking the circle of radius R centred at
the origin. If we write $\mathbf{r}=x+iy$ for the position vector in the
plane transverse to the magnetic field then the exact solution expressed in
Cartesian coordinates is given by $^{[18]}$ 
\begin{equation}
\mathbf{r=}\sqrt{\frac{J_a}{\pi m\omega _r}}exp[i\frac{(\omega _c+\omega
_r)\,t}2+\delta _a]+\sqrt{\left( \frac{J_R}{\pi m\omega _r}\right) }exp[i%
\frac{(\omega _c-\omega _r)\,t}2+\delta _A]  \tag{2.4}
\end{equation}
where $\omega _r=(\omega _c^2-2\omega _z^2)^{1/2},$ $(\omega _c+\omega
_r)/2\simeq \omega _c+\omega _z^2/2\omega _c$ and $(\omega _c-\omega
_r)/2\simeq \omega _z^2/2\omega _c.$ These results can evidently be
understood in terms of the azimuthal drift velocity 
\[
v_\phi =\frac{cE_r}{B_z}=\frac{\omega _z^2}{2\omega _c}r 
\]
which is a feature of charged particle motion in crossed electric and
magnetic fields. Expressed in the language of special relativity, the
electric field vanishes in a frame of reference rotating about the z-axis
with angular velocity $\omega _z^2/2\omega _c,$ and the particle behaves as
it would in a homogeneous magnetic field.

It needs to be borne in mind that the sudden lowering of the confining
potential on the ''gate electrode'' facing the detector during the release
phase is a non-adiabatic process whose influence on the trapped particles
may require further exploration. Since this failure of adiabaticity
corresponds to an injection of heat into the system there remains the
possibility that particles trapped in the vicinity of the gate may be lifted
into orbits such that they are lost on the electrode itself.\medskip 

\textbf{\#3 First Experiments at AERE\ Harwell 1970-75}

The apparatus, which is shown in Figure 1, was initially designed to operate
in magnetic fields up to 5T and accelerating potentials up to 50 kV. Some
unexpected problems were encountered in attempting to achieve these
conditions of operation. In the first experiments with the trapping system
it was found that:

\begin{figure}[!htb]
\centering
\includegraphics[width=0.8\textwidth]{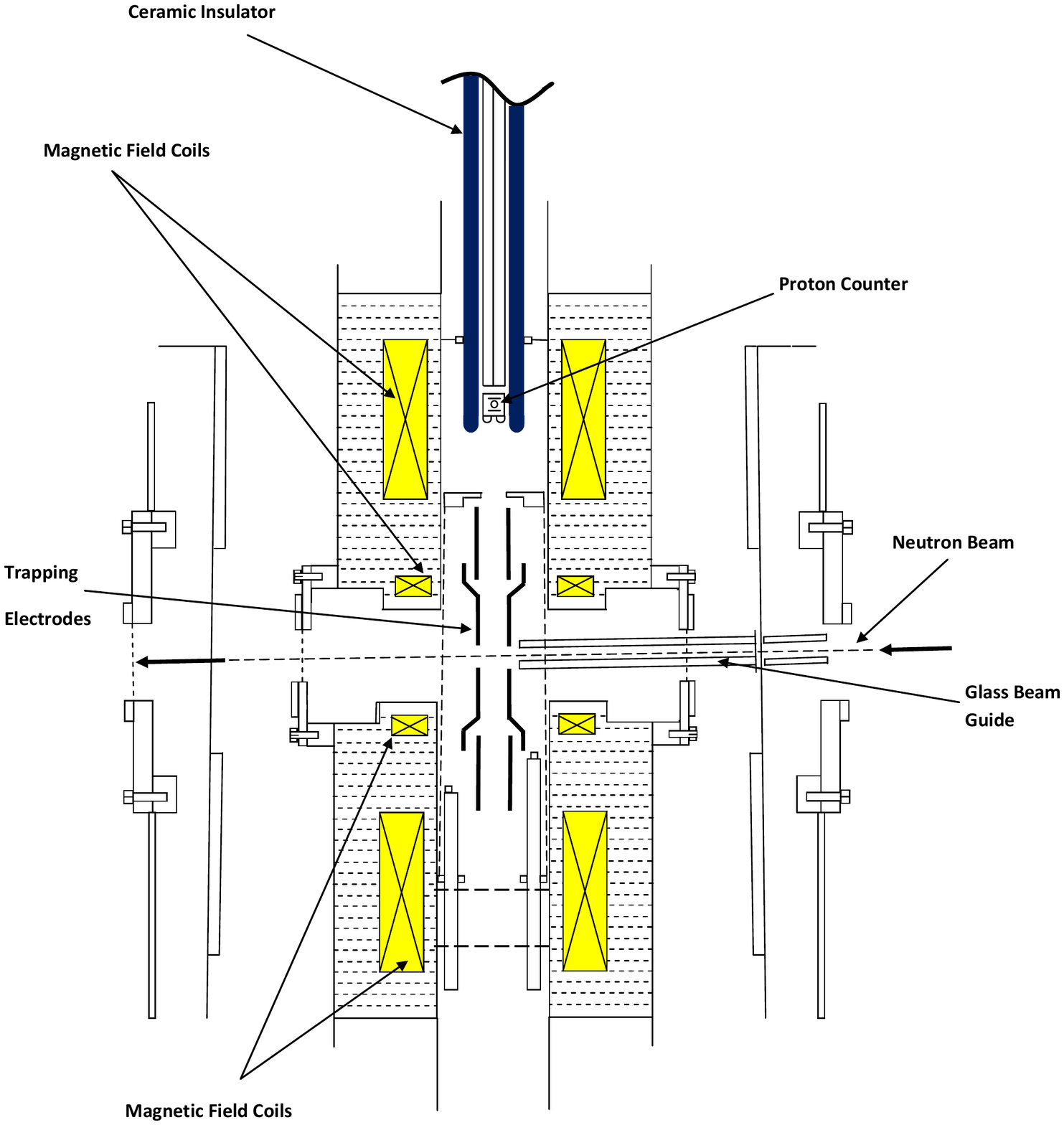}
\caption{The Penning trap used to detect low energy protons from
neutron $\beta $-decay at AERE, Harwell [Ref 21], and subsequently at the
Institut Laue-langevin, Grenoble [Ref 11]. The ceramic insulator was added
to the original design in order to protect against vacuum breakdown due to
the magnetron effect in the presence of crossed electric and magnetic
fields. The protons were detected in a silicon surface barrier detector, a
technique introduced here for the first time in the study of neutron decay,
which today is standard practice [Refs 30-32]} \label{figure}
\end{figure}

(i) The superconducting magnet would operate in persistent mode without
danger of quenching at currents $\leq 32$ amps in all four coils. In order
to reach higher currents, progressively longer run-up times were required
(1.5-2 hours), otherwise the magnet would go normal with immediate loss of
the liquid helium charge. In practice all four coils would run without
difficulty at 30 amps corresponding to magnetic fields of 1.6 T in the
centre of the trap and 4.0 T at the detector.

(ii) At zero magnetic field the accelerating electric field could be safely
raised to $\simeq -40$ kV; at higher potentials transitory breakdown pulses
occurred with increasing frequency.

(iii) With maximum magnetic field electrical breakdown was immediate and
total on application of 2-3 kV post-acceleration.

(iv) With an accelerating potential $\simeq -$30 kV electrical breakdown was
immediate and total when the magnet current reached 2-3 amps.

The most obvious feature of the observed pattern of breakdown in the
combined system of electric and magnetic fields was its very rapidity and
smoothness, quite unlike normal electrical breakdown whose onset is usually
preceded by periods of instability. Since elementary considerations would
indicate that breakdown transverse to a magnetic field should be hindered,
it was originally concluded that electrons, generated by some means or other
within the apparatus, were being chanelled along the magnetic field from the
detector to the upper ( ''pulsed'' or ''gate'') trapping electrode. However
a study of the breakdown charcteristics revealed that the dependence on the
electrode-detector separation was minimal and the discharge was taking place
in the space between the 2 cm diameter cylindrical tube containing the
detector signal and power leads and the 9.0 cm diameter cryostat wall which
is at ground potential.Thus the breakdown was associated with the presence
of crossed rather than parallel electric and magnetic fields.

The ultimate explanation for these observations appears to derive from the 
\textit{magnetron} effect, whereby electrons, generated in the annular gap
between the cylindrical tube carrying the detector leads and the
cyclindrical cryostat wall, move in cyclotron orbits which then precess at
right angles to both electric and magnetic fields. At sufficiently high
magnetic field this precessional motion is unimpeded and the electron orbits
carry out a free moton about the axis of the system. Ionization occurs in
the residual background gas producing more electrons which contribute to an
amplification of the process leading ultimately to breakdown following
electron diffusion to the cryostat wall.

The solution which has been successfully applied to this problem is to
enclose the detector tube in a coaxial beryllium oxide insulating cylinder
as shown in Figure 1. A procedure is then adopted whereby the high magnetic
field is established first and the high voltage is raised in steps of about
0.5-1.0 kV every few minutes.The reasoning behind this technique is that,
when the voltage is raised, the production of a single ion pair will be
followed by a mini-avalanche and the subsequent diffusion current will
deposit charge on the insulator rather than on the cryostat wall. Eventually
the point is reached where the electric field in the annular gap between
insulator and detector tube is reduced to a low value and avalanche
generation stops. When the discharge has ceased the potential is raised
again and the procedure is repeated until the final voltage is reached.

In the initial search for trapped protons from neutron decay the magnetic
field reached its minimum value of 1.6 T at the centre of the trap, rising
to $\simeq 4.0$ T, about 12.5 cms above the upper (pulsed ) electrode, and
again below the lower (mirror) electrode.This is a typical '\textit{magnetic
mirror}' configuration (see \textbf{\# 5}) although the significance of this
point was not fully appreciated at the time.This feature revealed itself in
the observation of a large number of magnetically trapped decay \textit{%
electrons }with an intensity essentially independent of the accelerating
voltage. It was therefore necessary to re-configure the magnetic field
profile such that the field decreased unifomly from the detector, through
the trap and beyond, thereby eliminating the magnetic mirror effect$^{\text{ 
}[21]}$. The magnetic field in the trap in the re-configured system attained
a value of 1.2 T.

It should also be pointed out that decay electrons of energy $<$ 1 keV\ can
be stored in the space inside the mirror electrode which,
while providing a potential barrier for protons,
 is a potential well
for electrons, which can generate background protons by ionization of
residual hydrogen. This effect may be identified from the non-statistical
rate of arrival of the spurious protons and is eliminated by reducing the
potential on the mirror to zero and resetting the trap before beginning each
trapping cycle. \medskip\ 

\textbf{\#4. Vacuum Breakdown in Crossed Electric and Magnetic Fields}

This is a phenomenon which has been explored experimentally in the greatest
detail by Penning$^{\text{ }[22]}$ .In the specific case of current interest
we consider the motion of an electron of mass $m_e$ and charge $-e$ moving
under the action of a uniform magnetic field B$_z$ in the cylindrical
annulus between a cathode of radius $r_i$ fixed at a potential $-V_0,$ and
an anode of radius $r_o>r_i$ .The electrostatic potential at radius $r$ is
then given by 
\begin{equation}
V(r)=-V_0\left( \frac{ln(r_o/r)}{ln(r_o/r_i)}\right)  \tag{4.1}
\end{equation}
We shall assume that the electron was initially emitted from the cathode
with zero kinetic energy so that $\dot{r}=r^2\dot{\phi}=0$ when $r=r_i.$
Since the conditions of cylindrical symmetry still apply we retain the
conservation of augular momentum about the axis 
\begin{equation}
p_\phi ==m_er^2(\dot{\phi}+\frac 12\omega _{ce})=m_er_i^2\left( \frac
12\omega _{ce}\right)  \tag{4.2}
\end{equation}
where $\omega _{ce}=(eB_z/m_e)$ has a positive value for an electron. The
total energy equation is then 
\begin{equation}
\mathcal{E}_e=\frac 12m_e\dot{r}^2+\frac 12m_e(\frac{\omega _{ce}}{2r}%
)^2\left( r^2-\text{ }r_i^2\right) ^2-e\left( V(r)-V_0\right)  \tag{4.3}
\end{equation}
\medskip\ The magnetron effect is initiated at that potential at which the
electron turns back, i.e. $\dot{r}=0$ at, say, $r=r_m<r_o$ and at lower
potentials the cyclotron orbits can precess freely about the axis thereby
generating avalanches in the background gas. At this point 
\begin{equation}
e\left( V(r_m)-V_0\right) =\frac 12m_e\omega _{ce}^2\left[ \frac{\left(
r_o^2-\text{ }r_i^2\right) ^2}{4r_o^2}\right] =\frac 12m_e\omega _{ce}^2a^2 
\tag{4.4}
\end{equation}
where $a$ is the cyclotron radius of electron motion in the magnetic field B$%
_z.$ In the original neutron lifetime experiments $r_i$=0.01m\ and $%
r_o=0.045\ $m$,$ thus $a$ = 0.0214m and B$_za=$0.107 Tm. From tabulated
values of the B$\rho $-parameters in electron spectroscopy we may conclude
that the electron energy at the magnetron transition point has a value close
to 3 MeV whch is well into the region of relativistic energies.We may also
invert the question and ask at what value of B$_z$ does the magnetron
transition take place when $V_0=30$ $kV$ ?. For $a$ = 0.0214m the answer is B%
$_z$= 0.028 T=280 gauss. The precise value is not important since it is
clear that the potential $V_0\simeq 2kV$ above which breakdown was observed
was not related to the magnetron effect which is already in full operation,
but rather to the electron energy required to generate an avalanche which in
the case of the proportional counter is typically of the order of few keV.

The charge to mass ratio for the electron has the value 
\[
\left( \frac{|e|}{m_e}\right) =1.758796\ \cdot 10^{11}C(kG)^{-1} 
\]
Thus the angular frequency of non-relativistic cyclotron motion in a 5 T
magnetic field is

\[
\omega _c=8.794\cdot 10^{11}\ sec^{-1} 
\]
and the radial electric field in the annulus is

\[
E_r(r)=-\left( \frac{\partial V}{\partial r}\right) =-V_0/(r\cdot
ln(r_o/r_i)) 
\]
which , for $V_0=30\ kV,$ gives the value 
\[
E_r(r)_{max}=-1.995\cdot 10^6\ Vm^{-1} 
\]
In order that the electron motion be oscillatory in the annular region it is
required that 
\begin{equation}
\left( \frac{\omega _r}2\right) ^2=\left( \frac{\omega _c}2\right) ^2+\left( 
\frac{eE_r(r)}{m_e}\right) >0  \tag{4.5}
\end{equation}
This requirement is easily satisfied since 
\[
\left( \frac{\omega _r}2\right) ^2=1.993\cdot 10^{23}-3.509\cdot 10^{19}\
sec^{-2}\gg 0 
\]

According to the Diethorn theory $^{[23]}$ of avalanche generation, these
occur under conditions of pressure $p$ where

\begin{equation}
V_0/pr_{i\text{ }}ln(r_o/r_i)>(2-10)\cdot 10^6(V/m)(bar)^{-1}  \tag{4.6}
\end{equation}
which at the observed breakdown point of $V_0\simeq 2kV$ indicates a
pressure range of 1.6-6.7 millibar.Application of Langevin theory $^{[24]}$
shows that, when \textbf{E} and \textbf{B} are orthogonal, the radial and
azimuthal drift velocities are given by

\begin{equation}
v_r=\left( \frac e{m_e}\right) \left[ \frac{\lambda _{coll}}{\lambda
_{coll}^2+\text{ }\omega _c^2}\right] E_{r\qquad }v_\phi =\left( \frac
e{m_e}\right) \left[ \frac{\omega _c}{\lambda _{coll}^2+\text{ }\omega _c^2}%
\right] E_{r\qquad }  \tag{4.7}
\end{equation}
where $\lambda _{coll}$ is the collision frequency between electrons and gas
molecules. For $\lambda _{coll}=0$, $v_r=0$ and $v_\phi =E_{r\text{ }}/B_z$,
which is the usual condition for motion in a vacuum. However, for $\lambda
_{coll}\neq 0,$ electrons will diffuse from cathode to anode and the system
breaks down. In practice the radial drift current can be cut off by
inserting a ceramic insulator in the annular region between cathode and
anode, as described in \textbf{\# 3,} and further breakdown is inhibited.

It is also important to understand that the fields may be crossed at the
point where the detector is positioned to record the accelerated protons,
putting the detector itself in danger of breakdown.This danger can be
avoided by recessing the detector approximately one diameter into its
containing tube, which is also at high negative voltage, where the radial
component of electric field approaches zero$^{\text{ }[14]}$. \medskip\ 

\textbf{\#5 The Magnetic Mirror Effect}

A proton of energy $E\approx 0.75$ keV moving in a magnetic field $B=5T$
carries out a cyclotron motion with angular frequency 
\[
\omega _c=4.80\cdot 10^8\sec ^{-1}\text{,} 
\]
on a circular orbit of radius $a<0.8$ mm. In a non-harmonic axial potential
the oscillation angular frequency is of course amplitude dependent, but,
approximating the potential on the axis by a quadratic dependence $V\left(
z\right) =V^{\prime \prime }\left( 0\right) z^2/2$, such that $V\left(
z\right) =1.0$ kV at $z=10$ cm, we may estimate 
\[
\omega _z=4.38\cdot 10^6\sec ^{-1}. 
\]
Thus $\omega _z$ is about 1\% of $\omega _c$, but is 100 times greater that
the angular frequency 
\[
\omega _p\simeq \omega _z^2/2\omega _c\approx 4\cdot 10^4\text{ }sec^{-1} 
\]
of the magnetron drift motion. Since the conditions required for the
conservation of the adiabatic invariants in the motion are then easily
fulfilled$^{\text{ }\left[ 20\right] }$, the longitudinal magnetic force
exerted on the trapped particle is then given by 
\begin{equation}
F_{mz}=-\frac{eB_r}r<r^2\dot{\phi}>=e\frac{\partial B_z}{\partial z}<r^2\dot{%
\phi}>=-m\omega _c\left( \frac 1{B_z}\frac{\partial B_z}{\partial z}\right)
<r^2\dot{\phi}>  \tag{5.1}
\end{equation}
where the factor 
\begin{equation}
<r^2\dot{\phi}>=\frac{J_a}{2\pi }=\frac{m\omega _c\,a^2}2+O\left[ (\frac{%
\omega _z}{\omega _c})^2\right]  \tag{5.2}
\end{equation}
is an adiabatic invariant of the motion.

Equation 5.1 defines the so-called `magnetic mirror' force which repels
charged particles from regions of high magnetic field irrespective of the
sign of the charge, and which has led to difficulties of interpretation in a
number of experiments on neutron beta-decay$^{\text{ }\left[ 25-26\right] }$%
. Essentially what happens is that the conservation of the longitudinal
adiabatic invariant $^{[20]}$ brings about a transfer of energy between the
transverse and longitudinal degrees of freedom, consistent with the
requirement that the magnetic force does no work. The magnetic force may
also be viewed as arising from a pseudo-potential 
\begin{equation}
\mathbf{F}_m=-\mathbf{\mu \cdot \nabla }B\mathbf{,\qquad }\mu =\frac
e{(-2\pi /\omega _c)}\pi a^2=(\frac m2)\text{ }a^2\omega _c^2/B  \tag{5.3}
\end{equation}
derived from the coupling of the magnetic moment $\mathbf{\mu }$ of the
cyclotron orbit and the magnetic field .

In the most recent version of the Penning trap method$^{\text{ }[13-14]}$
the magnetic field was designed to decrease by about 5\% between the trap
exit and the detector in order to ensure that the exiting protons were
impelled in the direction of the detector when the confining potential
barrier was lowered. In this way there was no possibility that protons
remained permanently trapped, thereby increasing the measured neutron
lifetime.

Subsequently it was noted that the magnetic mirror effect could be exploited
in reverse to measure the proton spectrum by setting the confining potential
barrier in a region where the magnetic field was only about 10\% of its
value at the centre of the trap, an arrangement which transfers most of the
proton's kinetic energy into its longitudinal degree of freedom $^{\left[
20\right] }$. This phenomenon, known as '\textit{adiabatic focusing}', also
provides the basis for the \textit{Fermi} process for the acceleration of
cosmic rays$^{\text{ }[27]}$ by a moving magnetic mirror.The process has
many features in common with the betatron accelerator.\medskip\ 

\textbf{\# 6 Radial Drifting}

For the study of neutron decay it is important to know that a proton which
is produced at a certain point in space, moving in a cyclotron orbit with
its guiding centre on a given magnetic field line, will move its guiding
centre onto an equivalent field line obtained by rotation through an
arbitrary angle about the $z$-axis. The concept of a guiding centre is valid
only when the motion is averaged over a period of time of order \TEXTsymbol{%
\vert}$2\pi /\omega _c$\TEXTsymbol{\vert}.If, however, there are substantial
departures from cylindrical symmetry, the guiding centre may end up on quite
a different quite field line having drifted away from the axis, and perhaps
out of the trap.The same phenomenon is of considerable significance in
plasma physics$^{[28]}$ .\ 

In the case of the Penning trap such azimuthal asymmetries may come about
mainly by

(a) a misalignment of electric and magnetic fields:

(b) an intrinsic asymmetry in the electric field due to slight deformation
of the electrodes into an elliptical shape:

(c) an intrinsic asymmetry in the magnetic field arising from asymmetric
coil winding.

In the case (a) of a misalignment the radial velocity is given by 
\begin{equation}
\left( \frac{dR(\mathbf{E})}{dt}\right) _1=-\left( \frac{\mathbf{n}\text{ }%
\times \text{ }\mathbf{E}}{B_z}\right) _r  \tag{6.1}
\end{equation}
where $\mathbf{n}$ is a unit vector in the direction of the magnetic field $%
\mathbf{B}$ which we assume to be cylindrically symmetric about the $z$%
-axis, whereas $\mathbf{E}$ is cylindrically symmetric about a $z^{\prime }$%
-axis which is set at a \textit{small} angle $\theta $ with respect to the $%
z $-axis.Thus we have the relations 
\[
x^{\prime }=xcos(\theta )+zsin(\theta ),\quad y^{\prime }=y,\quad z^{\prime
}=-xsin(\theta )+zcos(\theta ) 
\]
Assuming that the electric field in the $z^{\prime }$-system is derived from
a first order potential

\[
V^{(1)}(\mathbf{r}^{\prime })=-\left( \frac{m\omega _z^2}e\right) [\left(
\frac 12\right) r^{\prime \text{ }2}-z^{\prime }{}^{\text{ }2}]\approx
-\left( \frac{m\omega _z^2}e\right) [\left( \frac 12\right)
r^2-z^2+3rz\theta cos(\phi )] 
\]
it follows that 
\[
E_\phi =-\left( \frac 1r\right) \left( \frac{\partial V}{\partial \phi }%
\right) \approx \left( \frac{m\omega _z^2}e\right) 3z\theta sin(\phi ) 
\]

and 
\begin{equation}
\left( \frac{dR(\mathbf{E})}{dt}\right) _1^{(1)}=-(\frac{\omega _z^2}{%
2\omega _c})6z\theta sin(\phi )=-\omega _p6z\theta sin(\phi )  \tag{6.2}
\end{equation}
Assuming that $\omega _z\approx 10^7sec^{-1}$ and $\omega _p\approx
10^5sec^{-1},$and therefore $\omega _p/\omega _z\approx 10^{-2}$, the angle $%
\phi $ changes by an amount of order 1\% in half a precession of $z,$after
which time the drift velocity changes sign. Therefore the maximum total
drift is given when $\phi =\pi /2$ and $z$ goes from -$Z$ to +Z 
\[
\Delta R=-\omega _p6\theta \int_{-Z}^Zzdz=12\theta Z\omega _p/\omega _z 
\]
Assuming $\theta =1\%$ and Z=3 cm, this yields the value $\Delta R\approx 4%
\mathbf{\cdot }10^{-2}mm$ which is negligible.Also , since this drift
velocity changes sign every 10$^{-7}$seconds it is impossible for a
substantial drift to build up.

We may repeat the calculation taking into account the second order
correction to the potential

\[
V^{(2)}(\mathbf{r}^{\prime })\approx -\theta ^2\left( \frac{m\omega _z^2}%
e\right) [\left( \frac{-3}2\right) (x^2-z^2)] 
\]
from which, by a similar procedure, we may derive the result 
\begin{equation}
\left( \frac{dR(\mathbf{E})}{dt}\right) _1^{(2)}=-3\theta ^2\omega
_prsin(2\phi )  \tag{6.3}
\end{equation}
This equation may now be integrated to give 
\[
R(\phi )=R_{0\text{ }}exp[(3\theta ^2/2)cos(2\phi )] 
\]
Since the maximum value of $cos(2\phi )$ is unity and the minimum value is
zero the maximum radial displacement is 
\[
\Delta R=|R-R_0|=\left( \frac 32\right) \theta ^2r_{min\text{ }}\approx
10^{-3}mm,\quad \theta =1\%\quad R_0=5mm 
\]
We conclude that a small misalignment of the electric field produces minimal
radial drift.

A second possibility for finding a non-zero value for $E_\phi $ is a slight
deformation of the electrode into an elliptical shape.A potential which is
constant on the elliptical boundary 
\[
\frac{x^2}{\rho ^2(1-\varepsilon )^2}+\frac{y^2}{\rho ^2}=1 
\]
is obtained by adding a term of the form 
\[
V_\varepsilon (\mathbf{r})=-\varepsilon \left( \frac{m\omega _z^2}e\right)
[r^2cos^2(\phi )-z^2)] 
\]
so that the total potential is 
\[
V_{el}(\mathbf{r})^{}=-\left( \frac{m\omega _z^2}e\right) [\frac
12r^2(1+2\varepsilon \text{ }cos^2(\phi ))-z^2(1+\varepsilon \text{ })] 
\]
This potential satisfies Laplace's equation $\nabla ^2V_{el}(\mathbf{r})=0$
and is constant on the elliptical electrode.This corresponds to a relative
deformation in the radius $\rho $ of order $\varepsilon .$ It may be noted
that $V_{el}(\mathbf{r})$ is identical in form to $V^{(2)}(\mathbf{r}%
^{\prime })$, except that $\varepsilon $ replaces 3$\theta ^2/2$, i.e. 
\[
\left( \frac{dR(\mathbf{E})}{dt}\right) _2^{}=-2\varepsilon \omega
_pRsin(2\phi ) 
\]
an equation which can be integrated as before. Assuming that $\varepsilon
\approx 10^{-3}$ it follows that 
\[
\Delta R\approx 10^{-2}mm 
\]
and the deformation term is an order of magnitude larger that the second
order contribution due to a misalignment of electric and magnetic fields.

There are, in addition, two terms which describe radial drifting in a
cylindrically asymmetric \textbf{B}-field. These are 
\begin{equation}
\left( \frac{dR(\mathbf{B})}{dt}\right) _1^{}=\left( \frac{\mathbf{n}\text{ }%
\times \text{ (}\mu /e)\nabla B}B\right) _r  \tag{6.4}
\end{equation}
and 
\begin{equation}
\left( \frac{dR(\mathbf{B})}{dt}\right) _2^{}=\left( \frac{\mathbf{n}\text{ }%
\times \text{(p}_l^2/em)\partial \mathbf{n/}\partial s}B\right) _r  \tag{6.5}
\end{equation}
Here $p_{l\text{ }}$ is the component of momentum parallel to \textbf{B}, $%
p_{_{tr}\text{ }}$is the transverse component and 
\[
\mu =p_{tr\text{ }}^2/\left( 2mB\right) =(\frac m2)\text{ }a^2\omega _c^2/B 
\]
is the magnetic moment of the cycloton orbit (see eqn.5.3)

The magnetic field in the trapping volume is designed to be uniform only to
within about 1\% and we may assume that on the axis 
\[
B_z(z,0)=B_0[1+\alpha \left( \frac zZ\right) ^2] 
\]
where \TEXTsymbol{\vert}$\alpha |\approx 10^{-2}$. Therefore off-axis we
have the results

\[
B_z(z,r)=B_z(z,0)-(\frac r2)^2B_z^{^{\prime \prime }}(z,0)+...=B_0[1+\alpha
\{(\frac zZ)^2-(1/2)(\frac rZ)^2\}+..] 
\]

\[
B_r(z,r)=-rB_z^{^{\prime }}(z,0)+...=-2\alpha B_0(\frac rZ)(\frac zZ)+... 
\]
The first magnetic radial drift velocity is then given by

\begin{eqnarray*}
\left( \frac{dR(\mathbf{B})}{dt}\right) _1^{} &=&(\frac \mu {Be})(\frac 1R)%
\frac{\partial B}{\partial \phi }=(\frac 1R)[\left( \frac{B_r}{B_z}\right) 
\frac{\partial B_r}{\partial \phi }+\left( \frac{B_\phi }{B_z}\right) \frac{%
\partial B_\phi }{\partial \phi }+\frac{\partial B_z}{\partial \phi }] \\
&&
\end{eqnarray*}
where B$_\phi $ is zero and a $\phi $- dependent term has yet to be
introduced.The simplest winding error of order $\eta $ would have to take
the form $cos(2\phi )$ in the transverse plane and this could be described
by a scalar magnetic potential, similar to$V_\varepsilon (\mathbf{r})$,
which satisfies Laplaces equation. 
\[
\Psi _\eta (r)=-\eta (\frac{B_0}Z)(r^2cos^2(\phi )-z^2) 
\]
We then find after some computation

\[
\left( \frac{dR(\mathbf{B})}{dt}\right) _1=(\frac \mu {Be})(\frac 1R)\frac{%
\partial B}{\partial \phi }=(\frac \mu {Be})\left( (\frac{-\eta ^2}{B_z})(%
\frac{B_0}Z)^22Rsin(2\phi \right) \approx (\frac \mu {Be})(\frac{\eta ^{}}%
Z)^2B_0R\frac{dcos(2\phi )}{d\phi } 
\]

Writing 
\[
\frac{dcos(2\phi )}{d\phi }=\frac{dcos(2\phi )}{dt}/\frac{d\phi }{dt}=\frac{%
dcos(2\phi )}{dt}/\omega _p 
\]
the radial drift equation can now be integrated to give 
\begin{eqnarray*}
R &=&R_0exp\left( (\frac{ma^2\omega _c^2}{2mc^2})(\frac{c^2}{\omega _c\omega
_p})\frac{\eta ^2}{Z^2}cos(2\phi )\right) \\
&&
\end{eqnarray*}
The maximum drift then occurs when $cos(2\phi )=1.$Also, since for
neutron-decay protons $ma^2\omega _c^2$/$2mc^2$ \TEXTsymbol{<}10$^{-6}$, it
follows that for $\omega _c=5\mathbf{\cdot }10^8$, $\omega _p=10^5$ and Z=3
cm, $R/R_0\approx 2\eta ^2$. Since the field is designed to be uniform to
within 1\% we may assume that $\eta $ $\ll 0.1\%$ and $\Delta R\ll 10^{-5}mm$
$.$

The second magnetic drift velocity is given by 
\begin{equation}
\left( \frac{dR(\mathbf{B})}{dt}\right) _2^{}=\left( \frac{m\dot{z}^2}{eB}%
\right) \{\left( \frac{B_\phi }B\right) \frac{\partial B_z}{\partial s}%
\left( \frac{B_z}B\right) -\left( \frac{B_z}B\right) \frac \partial
{\partial s}\left( \frac{B_\phi }B\right) \}  \tag{6.6}
\end{equation}

\[
\approx \left( \frac{m\dot{z}^2}{eB_0}\right) \left( \frac{\eta Rsin(2\phi )}%
Z\right) \left( \frac{2\alpha \text{ }z}{Z^2}-\frac{2\eta }Z\right) 
\]
The term proportional to $2\alpha $ $z$/$Z^2$changes sign every period of
z-oscillation and may therefore be ignored. We then find that 
\[
\left( \frac{dln(R(\mathbf{B}))}{dt}\right) _2^{}=\left( \frac{m\dot{z}^2}{%
mc^2}\right) \left( \frac{c^2}{\omega _c}\right) \left( \frac \eta Z\right)
^2\frac d{dz}(cos(2\phi ) 
\]
As before this equation can be integrated to give 
\[
R=R_0exp(\left( \frac 12\frac{m\dot{z}^2}{mc^2}\right) \left( \frac{2c^2}{%
\omega _c\omega _p}\right) \left( \frac \eta Z\right) ^2cos(2\phi ) 
\]
Apart from the additional factor of 2 this result implies that the two
magnetic drift terms are about the same and equally negligible. Of course
this is not true in the case that the magnetic field lines are deliberately
designed to bend $^{[12-14]}$. \medskip

\textbf{\#7 Proton Loss by Transverse Diffusion }.

Charged particles contained in a Penning trap in conditions of perfect
vacuum will stay trapped forever. Unfortunately perfect vacuum cannot be
achieved in practice and the trapped particles will undergo collisions with
atoms of residual gas. In the case of a cryo-pumped system helium atoms are
likely to be the most important scattering centres. As a result the guiding
centres of the cyclotron orbits will suffer random displacements. Since
there is an applied electrostatic field in the form of a\textit{\
longitudinal} potential well, the protons are prohibited by energy
conservation from escape \textit{along} the magnetic field whether or not
they undergo collisions. However, since there is an outward directed radial
electric field in the trap these protons can be transported by successive
collisions \textit{transverse} to the magnetic field lines and must
eventually be lost on the electrode walls.

To make further progress it is necessary to inquire into the details of the
individual collision processes. There are two extreme situations
corresponding to the conventional classification of collisions into close
and distant encounters (i) In a close encounter the guiding centre may be
displaced through the maximum amount, equal to twice the radius of gyration,
correponding to Poissonian modulation of the free motion.(ii) In a distant
collision the displacement is infinitesimal for a single collision but,
since the number of such collisions is large the net displacement is
finite.This situation may be described as Gaussian modulation of the free
motion, and seems likely to dominate assuming that individual collisions
between protons and residual atoms are governed by a shielded Coulomb
potential, whose differential cross section 
\begin{equation}
\frac{d\sigma (\theta )}{d\Omega }=\frac 1{16}\left( \frac{Ze^2}{4\pi
\varepsilon _0\mathcal{E}_p}\right) ^2[sin^2\theta +\eta ^2]^{-2}  \tag{7.1}
\end{equation}
is strongly peaked in the forward direction. Here $\eta =\hbar /2r_s$p$_p$
where r$_{s\text{ }}$is the shielding radius. This latter is a somewhat
uncertain quantity but for rare gas atoms is typically of the order of half
of one Bohr radius and therefore $<$ 0.5 \.{A}. Assuming that the mean free
path between collisions is small in comparison with the radius of gyration,
then the mean \textit{density} of trapped protons $F(r,t)$ at radius $r$ at
time $t$ satisfies the diffusion equation 
\begin{equation}
\frac{\partial F(r,t)}{\partial t}=D\nabla ^2F(r,t)  \tag{7.2}
\end{equation}
where 
\begin{equation}
D=(\nu /2)\cdot <x^2>  \tag{7.3}
\end{equation}
is the diffusion coefficient, $\nu $ is the collision rate per unit time and 
$<x^2>$ is the mean square displacement per collision \textit{transverse} to
the magnetic field. $F(r,t)$ is subject to the spatial boundary condition $%
F(r,t)=0$ when $r=r_e$ where $r_e$ is the electrode radius.

The diffusion equation has to be solved subject to a second boundary
condition which specifies the density $F(r,0)$ of trapped particles at zero
time.The solution of this equation is quite lengthy $^{[29]}$ and leads to
the result that, if the trap is filled at a uniform rate n$_0/\tau $, where $%
\tau $ is the trapping time, the number of trapped protons at time $\tau $
is given by

\begin{equation}
n(\tau )=\int_0^{r_e}F(r,\tau )2\pi rdr=4n_0\sum_{m=0}^\infty \frac{%
rJ_1\left( \alpha _m\frac r{r_e}\right) }{r_eJ_1(\alpha _m)}\left( \alpha
_m^4D\tau /r_{e^{}}^2\right) ^{-1}\left[ 1-exp(\alpha _m^2D\tau
/r_e^2)\right]  \tag{7.4}
\end{equation}
For scattering in the centre of mass frame of protons of energy $\mathcal{E}%
_p$ and momentum p$_{p\text{ }}$on residual atoms of atomic number Z ,via a
shielded Coulomb potential, the diffusion coefficient is found to be

\begin{equation}
D=\frac \pi 6\left( \frac{Ze^2}{4\pi \varepsilon _0\mathcal{E}_p}\right)
^2Nva^2\left( ln[1+1/\eta ^2]-[1+1/\eta ^2]\right)  \tag{7.5}
\end{equation}
where N is the number density of residual atoms and $v$ is the proton
velocity.. Assuming a background pressure of 10$^{-8}torr$, coresponding to
a number density of helium atoms of 3.5$\cdot $ 10$^{14}m^{-3}$ it has been
estimated $^{[12]}$ that the root mean square position of the guiding centre
drifts by about 1 mm in 180 seconds which means that for trapping times $%
\tau \leq 10$ $ms$ $^{[12-14]}$ proton loss by diffusion across the magnetic
field lines is negligible.\medskip\ 

\begin{center}
\textbf{REFERENCES}$^{}$
\end{center}

\begin{enumerate}
\item  \textbf{[}1\textbf{] }Bahcall J.N. and May R.M., Astrophys.J.\textbf{%
155 }(1969) 502

[2] Bahcall J.N., Phys. Lett. B \textbf{338} (1994) 276.

[3] Dubbers D. Mampe W. and Doehner J., Europhys. Lett. \textbf{11} (1990)
195.

[4] Dubbers D., Prog. Part. Nucl. Phys. \textbf{26} (1991) 173.

[5] Byrne J., Proc. Nobel Symposium \textbf{91, (}World Scientific
Singapore1995\textbf{)} 311

[6] Hardy J.C. and Towner I.S., Phys.Rev. C \textbf{71 }(2005) 055501;
Phys.Rev.Lett.\textbf{94} (2005) 092502

[7] Paul W. et al., Z. Phys. C \textbf{45} (1989) 25.

[8] Huffman P.R. et al., Nature \textbf{403} (2001) 62

[9] Mampe W. et al., Phys. Rev. Lett. \textbf{63} (1989) 593.

[10] Serebrov A.P. et al., Phys.Rev. C \textbf{78} (2008) 03550

[11] Byrne J. et al., Phys Lett. B \textbf{92} (1980) 274

[12] Williams A.P., D.Phil Thesis, University of Sussex (1989) pages 36-8

[13] Byrne J .et al., Phys. Rev. Lett. \textbf{68} (1990) 289; Europhys.
Lett. \textbf{33} (1996)187

[14] Byrne J.et al., J. Phys. G. Nucl. Part. Phys. \textbf{28} (2002) 1325

[15] Penning F.M., Physica \textbf{4} (1937) 71

[16] Dehmelt H.G., Adv. At. Mol. Phys. \textbf{3} (1967) 53; \textbf{5}
(1969) 109

[17] Werth G., J. Phys.G Nucl. Part. Phys.\textbf{20} (1994) 1865.

[18] Byrne J and Farago P.S., Proc. Phys. Soc. \textbf{86} (1965) 801

[19] Byrne J and Dawber P.G., Nucl. Instr .and Meth A \textbf{332} (1993)
363; A \textbf{349} (1994) 454;
AIP Conf. Proc. 457 (AIP  Woodbury New York 1999) page 103

[20] Byrne J., Proc. Roy. Soc. Edin.\textbf{70 }(1972) 47

[21] Byrne J., Inst. Phys. Conf. Ser. No \textbf{42}
(1978) page 28;  No \textbf{64}  (1983) page 15;  (IOP Bristol and London) 

[22] Penning F.M., Physica \textbf{3} (1936) 873

[23] Knoll G.F., \textit{Radiation Detection and Measurement}, (John Wiley
1979), page 193.

[24] Blum W. and Rolandi L.,\textit{\ Particle Detection with Drift Chambers}
(Springer-Verlag 1994) page 53..

[25] Bopp B.et al., Phys. Rev. Lett. \textbf{56} (1986) 919.

[26] Klempt E. et al., Phys.Rev. C \textbf{37} (1988) 179.

[27] Fermi E., Phys.Rev.\textbf{75} (1949)1169; Astrophys J.\textbf{119}
(1954)1

[28] Chandrashekar S., \textit{Plasma Physics}, (University of Chigago
Press1960), page 65

[29] Byrne J., CRYSAB \textbf{1 }(1975) 33

[30] Baessler S. et al., Europhys.J., A \textbf{38} (2008)17

[31] Wietfeldt F.E. et al., Nucl. Instr. and Meth A \textbf{611 }(2009) 207

[32] Pocanic D. et al., Nucl. Instr.and Meth A \textbf{611 }(2009) 211

\medskip\medskip 

\textbf{Acknowledgement}

I should like to express my gratitude to Ferenc Gl\"uck at the Karlsruhe
Institute of Technology for his advice, support and encouragement in the
preparation of this review.
\end{enumerate}

\end{document}